\documentclass[prl,twocolumn,showpacs]{revtex4}
\usepackage{graphicx,amssymb,amsmath}
\begin{document}
\title{Inelastic Scattering from Local Vibrational Modes}
\author{Bal\'azs D\'ora}
\email{dora@pks.mpg.de}
\affiliation{Max-Planck-Institut f\"ur Physik Komplexer Systeme, N\"othnitzer Str. 38, 01187 Dresden, Germany}
\author{Mikl\'os Gul\'acsi}
\affiliation{Max-Planck-Institut f\"ur Physik Komplexer Systeme, N\"othnitzer Str. 38, 01187 Dresden, Germany}

\date{\today}

\begin{abstract}
We study a nonuniversal contribution to the dephasing rate of conduction electrons due to local vibrational modes. 
%Bosonization allows us to evaluate the full T-matrix.
The inelastic scattering rate is strongly influenced by multiphonon excitations, exhibiting oscillatory behaviour. For 
higher frequencies, it saturates to a finite, coupling dependent value. 
In the strong coupling limit, the phonon is 
almost completely 
softened, and the inelastic cross section reaches its maximal value. 
This represents a magnetic field insensitive contribution to the dephasing time in mesoscopic 
systems, in addition to magnetic impurities.

\end{abstract}


\maketitle

%\section{Introduction}

The loss of quantum coherence, occurring through inelastic processes during which some excitations are left behind and 
the outgoing state is not the a single particle state any more, is characterized by the dephasing time, $\tau_\phi$.
The dephasing time can reliably be estimated from the low-field magnetoresistance data from the weak-localization
corrections to the Drude conductivity\cite{mohanty}.
In general, any dynamical impurity with internal degrees of freedom can dephase conduction electrons. If the 
impurity changes its state, the environment felt by the electrons changes, causing dephasing.

Recently, dephasing due to magnetic impurities has been addressed successfully in the relevant temperature range (around 
and below the Kondo temperature\cite{hewson}) by relating the dephasing time to the inelastic
cross section\cite{borda,micklitz,bauerle}.
The latter can be determined from the knowledge of the many-body $T$-matrix. It accounts nicely for 
various features of the dephasing time such as its saturation above the Kondo temperature.
However, in many systems, Coulomb-type interaction is not the only source of correlation,
which leads us to consider the interaction of electrons with local vibrational (phononic) modes\cite{galperin}.
The interest to study these model is at least threefold:
first, artificial quantum dots based on single molecules\cite{park}  (e.g. C$_{60}$) usually distort upon the addition 
or removal of 
electrons, and can act as a quantized nano-mechanical oscillator by 
single electron charging\cite{park}. Moreover, other realization contains suspended quantum dot cavity\cite{weig}, where 
confined phonon modes influence the transport.
Second, in many strongly correlated systems, such as heavy fermions or valence fluctuation systems, lattice
vibrations are known to couple strongly to electrons. The recent discovery of magnetically robust heavy fermion
behaviour in filled skutterudite compound\cite{sanada} SmOs$_4$Sb$_{12}$ renewed interest in Kondo phenomena with
phononic
origin\cite{yotsuhashi,hotta}. Both the specific heat coefficient and coefficient of the quadratic temperature
dependence of the electrical resistivity were found to be almost independent of an applied magnetic field.

Finally, yet another motivation arises from recent experiments on Cu$_{93}$Ge$_4$Au$_3$ thin films\cite{huang}, 
where the observed dephasing time was found to be rather insensitive to high magnetic fields, suggesting non-magnetic 
dephasing processes, such as dynamical defects.

\begin{figure}[h!]
\centering
\centering{\includegraphics[width=6cm,height=4cm]{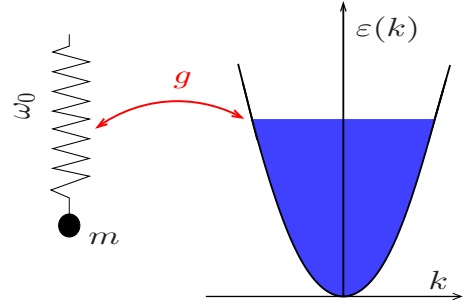}}
\caption{(Color online) Schematic picture of a nanomechanical resonator coupled to a conduction electron bath. 
\label{rugoelektron}}
\end{figure}

Therefore, it is timely to investigate other source of dephasing, than magnetic impurities, 
such as a localized oscillator\cite{galperin}.  
In general, these dynamic point defects are randomly distributed in a  system. 
By assuming that the impurities are distributed sufficiently far from each other, it suffices to study one single mode, 
with fixed energy $\omega_0$ (Einstein oscillator), similarly to the treatment of magnetic impurities\cite{hewson}.
The phonon can couple to the local charge density, and modify the local potential felt by the electron (single 
impurity Holstein model\cite{fransson,cornaglia1,schwabe,kral}). 
In the most general case, this amounts to study the Yu-Anderson or single impurity Holstein model (visualized in 
Fig. \ref{rugoelektron}) given 
by\cite{gadzuk,yuanderson,irkhin}
\begin{equation}
H=\sum_{k}\varepsilon(k)c^+_{k}c_{k}+g_{d}Q\Psi^+({\bf 0})\Psi({\bf 0})+\frac{P^2}{2m}+\frac 
{m\omega_0^2}{2}Q^2,
\label{hamilton}
\end{equation}
which describes 3-dimensional electrons interacting with a local bosonic mode, 
$\Psi({\bf r})$ denotes the 
bulk 3D 
electron field operator and $c_{k}$ its Fourier components.
%, $n=1..N$ denotes the (orbital or spin) degeneracy of the electron states ($N=2$ for spin). 
%Since scattering is the most relevant 
%at the Fermi energy, we neglect the wavevector dependence of $g_{k,k^\prime}$ in the %followings\cite{doraphonon,fransson}. 
This accounts for intrinsic and 
extrinsic point defects such as interstitial or substitutional  impurities\cite{montroll}. From the latter class, 
foreign 
atoms (like Fe in Ag or Au\cite{mohanty,bauerle}) mainly represent a source of potential scattering. Beyond this, 
they are capable of 
magnetic scattering\cite{borda,micklitz}, and can cause a modification of the local phononic environment as well, 
resulting in a 
Hamiltonian like Eq. \eqref{hamilton}.
Besides foreign atoms, the self-interstitial impurities (proper atom at a non-lattice site) also distort the 
local potential background felt by the conduction electrons, leading to local electron-phonon interaction.
This model has a rich history in the field of nonlinear impurities\cite{molina1}.
In quantum optics, such model arises in arrays of nonlinear optical waveguides for example.
Moreover, in the condensed matter context, Eq. \eqref{hamilton} bears close resemblance  to the Anderson-Holstein 
impurity model\cite{hewsonmeyer,cornaglia}, describing molecular vibrations in quantum dots.
It can be regarded as a generalization of scattering on two-level systems to the multi-level case\cite{zawadowski}.

Due to the isotropic nature of the electron-phonon coupling, the electron field operators can be expanded in 
the appropriate angular momentum channels (depending on the dimensionality of the conduction electrons), and 
only the s-wave component couples to the impurity\cite{nersesyan} as
\begin{gather}
\Psi({\bf r})=\frac{e^{ik_Fr}R(r)-e^{-ik_Fr}R(-r)}{2i\sqrt\pi r}+\begin{array}{c}
\textmd{higher}\\
\textmd{harmonics}
\end{array},
\label{3Dchiral}
\end{gather}
where $R(x)$ stands for the chiral, right moving radial component, $r\geq0$ is the radial coordinate, $k_F$ is the 
Fermi wavenumber.
Thus, the model can be mapped onto an effective model 
of 
chiral fermions, interacting with the bosonic mode at the origin, and is ideal for bosonization 
($R(x)=\exp[i\sqrt{4\pi}\Phi(x)]/\sqrt{2\pi\alpha}$):
\begin{gather}
 H=-iv\int\limits_{-\infty}^{\infty} dx
R^+(x)\partial_x R(x)+gQ\rho(0)+\frac{P^2}{2m}+\frac {m\omega_0^2}{2}Q^2,
 \label{hamilton1d}
\end{gather}
and only the radial motion of the particles is accounted for by chiral (right moving) fermion
field, $\rho(x)=:R^+(x)R(x):$ is the normal ordered electron charge density, $v$ is the 
Fermi 
velocity, $g=g_{d}k_F^2/\pi$ describes the local electron-phonon coupling, $m$ and $\omega_0$ are the phononic 
mass and frequency, respectively, $Q$ and $P$ are the phonon displacement field and momentum conjugate to 
it. 
In the spirit of the Yu-Anderson model\cite{yuanderson}, the phonons couple to the electronic density fluctuations
 in our model as well. The ground state will be degenerate, similarly to the Kondo 
problem\cite{hewson}, between 
occupied and unoccupied electron states, leading to the formation of a double well potential\cite{yuanderson}. The bosonized Hamiltonian is identical to that of coupled harmonic oscillators\cite{weiss,doraphonon}.

Due to the electron phonon coupling $g$, the phonon mode softens as\cite{gadzuk,doraphonon}
\begin{equation}
\omega_{p\pm}=-i\Gamma\pm\sqrt{\omega_0^2-\Gamma^2-\Gamma\omega_0^2/\Gamma_2},
\end{equation}
where $\Gamma_2=\pi\omega_0^2/4W\ll\omega_0\ll W$, $W$ is the bandwidth of the conduction electrons, and 
$\Gamma=\pi(g\rho)^2/2m$ for small $g$, and approaches $\Gamma_2$ as $g\rightarrow\infty$.
Here, $\rho=1/2\pi v_F$ is the chiral electron density of states.
The explicit dependence of $\Gamma$ on $g$ cannot be determined by the 
bosonization approach\cite{doraphonon}. The real part of the phonon frequency remains finite (underdamped) for 
$\Gamma<\Gamma_1\approx \Gamma_2(1-\Gamma_2^2/\omega_0^2)$. For 
$\Gamma_1<\Gamma<\Gamma_2$, 
the oscillatory behaviour disappears from the phononic response (Re$\omega_{p\pm}=0$), and two distinct 
dampings characterize it (overdamped). This softening is observable in many physical quantities, such as the enhancement 
of the local 
charge response, indicating the charge-Kondo phenomenon.

The local retarded Green's function of the conduction electrons is defined, using Eq. \eqref{3Dchiral} 
as\cite{kakashvili}
\begin{gather}
G_R(t)=-i\Theta(t)\langle\{\Psi(x=0,t),\Psi^+(y=0,t=0)\}\rangle=\nonumber\\
=-i\Theta(t)\frac{k_F^2}{4\pi}\sum_{\gamma,\gamma'=\pm}\lim_{\alpha\rightarrow 0}
\langle\{R(\gamma\alpha,t),R^+(\gamma'\alpha,0)\}\rangle=
\nonumber\\
=-i\Theta(t)\frac{k_F^2}{8\pi^2}\sum_{\gamma,\gamma'=\pm}\left(\exp[C_{\gamma,\gamma'}(t)]+
\exp[C_{\gamma',\gamma}(-t)]\right),
\label{green1}
\end{gather}
where $C_{\gamma,\gamma'}(t)=\lim_{\alpha\rightarrow 0}4\pi\langle\Phi(\gamma 
\alpha,t)\Phi(\gamma'\alpha,0)-[\Phi(\gamma\alpha,t)^2-\Phi(\gamma'\alpha,0)^2]/2\rangle-
\ln(\alpha)$, whose second appearance in 
Eq. \eqref{green1} follows from 
time reversal symmetry, and $\alpha\sim 1/W$ is the short distance cutoff in the bosonized theory, $\gamma$ and 
$\gamma'$ denotes the sign of the space coordinate. 
The impurity site is defined in terms of the chiral operators as the limiting value from Eq. \eqref{3Dchiral}. The 
$\gamma=\gamma'$ combinations does not contain any information about the presence 
of the impurity at the origin\cite{fuentes}, since both field operators are taken on the same side of the impurity, and 
no scattering 
is experienced. On the other hand, $\langle R(\gamma\alpha,t)R^+(-\gamma\alpha,0)\rangle$ 
describes scattering through the impurity, and depends strongly on its properties. 
%Eq. \eqref{green1} differs from the one studied in Ref. \onlinecite{irkhin,doraphonon}, and is more appropriate for 
%the present case.
The expectation value $C_{\gamma,\gamma'}(t)$ at 
bosonic Matsubara frequencies can be evaluated from the bosonized 
Hamiltonian\cite{doraphonon}. 
Then by analytical continuation to real frequencies, we can determine the 
spectral intensity following Ref. \onlinecite{zubarev}, which leads to the desired function, $C_{\gamma,\gamma'}(t)$. 
This follows the derivation of the position autocorrelator of a harmonic oscillator coupled to a heat 
bath\cite{weiss}. The local retarded
Green's function follows as
\begin{gather}
G_R(\omega)=-i\pi\frac{\rho_b}{2}\left(1+F(\omega)+\left(\dfrac{\omega_{p+}}{\omega_{p-}}\right)^{4\Gamma 
i/(\omega_{p+}-\omega_{p-})}\right),
\label{retgreen}
\end{gather}
where
\begin{gather}
F(\omega)=\int\limits_0^\infty {dt} \frac{\exp(i\omega t)}{\pi t}\textmd{Im}\exp
\left[\frac{4\Gamma (d[t\omega_{p+}]-d[t\omega_{p-}])}{i(\omega_{p+}-\omega_{p-})}\right]
\end{gather}
and
%\begin{gather}
$d[x]=\exp(-ix)\left(-\textmd{Ci}(-x)+i\left(\textmd{Si}(-x)-\frac\pi 2\right)\right)$
%\end{gather}
in the underdamped case, Si$(x)$ and Ci$(x)$ are the sine and cosine integrals, and 
$\rho_b=k_F^2/2\pi^2v_F$ is the 3D bulk density of states. When Eq. \eqref{retgreen} is expanded in $\Gamma$, 
it agrees with perturbative results\cite{engelsberg}.
We can identify the $T$-matrix using the Dyson equation for the local 
Green's function as
%\begin{equation}
$G(0,\omega)=G_0(0,\omega)+G_0(0,\omega)T(\omega)G_0(0,\omega),$
%\label{tmatrix}
%\end{equation}
where $G_0(0,\omega)=-i\pi\rho_b$ is the unperturbed Green's function.
Therefore, we determine the full T-matrix as
\begin{gather}
T(\omega)=\frac{i}{2\pi\rho_b}\left(\left(\dfrac{\omega_{p+}}{\omega_{p-}}\right)^{4\Gamma
i/(\omega_{p+}-\omega_{p-})}+F(\omega)-1\right)
\end{gather}
which is the main result of this paper. It contains all the non-perturbative effects of the quantum impurity introduced 
in Eq. 
\eqref{hamilton}.
As was shown in Ref. \onlinecite{borda}, this possesses all the information 
needed to evaluate the various cross sections.
\begin{figure}[h!]
\centering
%\psfrag{x}[t][b][1][0]{Re$s(\omega)$}
%\psfrag{y}[][][1][0]{Im$s(\omega)$}  
%\psfrag{o}[b][t][1][0]{$\omega=0$}
\centering{\includegraphics[width=7.6cm,height=7.6cm]{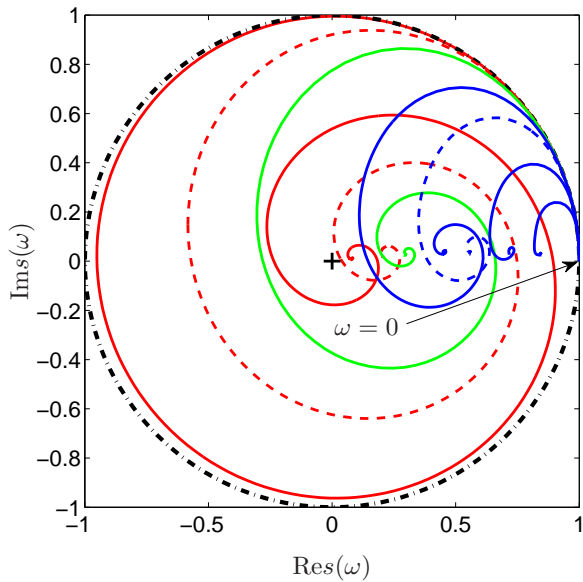}}

\caption{(Color online) 
The evolution of the $s$-matrix on the complex plane is shown at $T=0$ for $W=10\omega_0$,  $\Gamma/\Gamma_2=0.3$, 0.5, 0.7 (blue dashed), 0.8 (blue lines), 0.9 (green line), 0.94 (red dashed line) and 0.98 (red solid line) from right to left. 
Close to the critical coupling $\Gamma=\Gamma_2$, the scattering is dominantly elastic for small frequencies, and follows the unit circle. Then, it reaches rapidly the maximal inelastic rate ($s=0$) with increasing $\omega$, because at the critical coupling, the phonon excitation energy $|\omega_{p-}|\rightarrow 0$, facilitating inelastic processes.} 
\label{sigmao0}
\end{figure}
\begin{figure}[h!]
%\vspace*{5mm}
%\psfrag{x}[t][b][1][0]{$\omega/\omega_0$}
%\psfrag{y}[][][1][0]{$\sigma/\sigma_0$}   
%\psfrag{g1}[b][t][1][0]{$\Gamma/\Gamma_2=0.3$}
%\psfrag{g2}[b][t][1][0]{$\Gamma/\Gamma_2=0.6$}
%\psfrag{g3}[b][t][1][0]{$\Gamma/\Gamma_2=0.9$}
%\psfrag{g4}[b][t][1][0]{$\Gamma/\Gamma_2=0.99$}

\centering{\includegraphics[width=7.5cm,height=7.1cm]{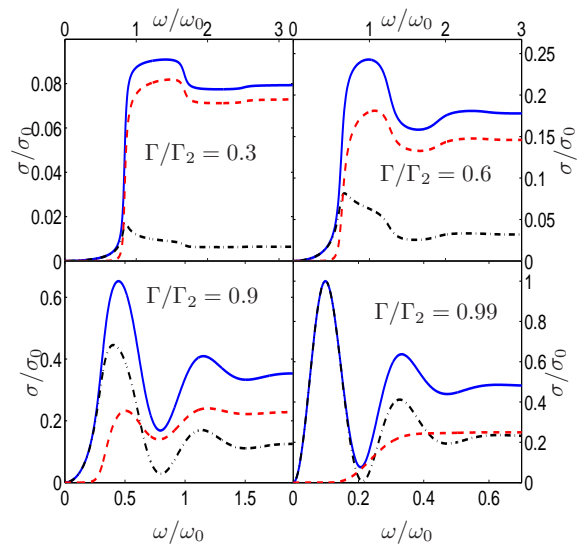}}

\caption{(Color online) The frequency dependent scattering cross sections are shown for $W=10\omega_0$, 
$\Gamma/\Gamma_2=0.3$, 0.6, 0.9 and 0.99 from left to right, top to bottom. The solid blue, red dashed and
black
dashed-dotted line denotes $\sigma_{tot}$, $\sigma_{inel}$ and $\sigma_{el}$, respectively.
For small frequencies, all scattering rates vanish. Then, wild oscillations show up, representing the various multiphonon ($n$-phonon) processes, setting in at $n\times|\omega_{p\pm}|$.
At higher frequencies ($|\omega_{p\pm}|\ll \omega \ll W$), inelastic processes dominate. With increasing $\Gamma$, the weight of both elastic and inelastic processes increase. By approaching the critical coupling, the inelastic scattering rate reaches its maximal value, $\sigma_0/4$, because with the vanishing phonon frequency, $\omega_{p-}$, phonon excitations become more likely, favouring inelastic scatterings. Note the different horizontal and vertical scales on the figures.}
\label{sigmaGall}
\end{figure}
From this, we obtain the many-body $s$-matrix as\cite{langreth}
\begin{equation}
s(\omega)=1-i2\pi\rho_b T(\omega),
\end{equation}
which stays always within the unit circle in the complex plane. The circumference of the unit circle denotes perfect 
elastic scattering.
The total, elastic and inelastic cross section is readily evaluated from $s(\omega)$ as\cite{langreth,borda,micklitz}
\begin{gather}
\sigma_{tot}=\frac{\sigma_0}{2}(1-\textmd{Re}s(\omega)),\\
\sigma_{inel}=\frac{\sigma_0}{4}(1-|s(\omega)|^2),\\
\sigma_{el}=\frac{\sigma_0}{4}(1-2\textmd{Re}s(\omega)+|s(\omega)|^2),
\end{gather}
where $\sigma_0=2/\pi\rho_b v$ is the maximal value of the cross section. 
$\sigma_{inel}$ is directly related to the dephasing time $\tau_{\phi}$, measured in 
weak-localization experiments\cite{borda}. 
A localized vibrational mode contributes to the inelastic scattering rate even at $T=0$.
The evolution of 
its $s$-matrix is shown in Fig. \ref{sigmao0}. In the 
weak coupling limit (small $\Gamma$), scattering is negligible. For small frequencies,
the $s$-matrix follows the upper semicircle, thus scattering is mostly elastic.
For larger $\Gamma$, the 
weight of the inelastic component increases with $\Gamma$ until it reaches 
its maximum value at $s=0$ at the critical coupling $\Gamma_2$. There, the total scattering cross 
section is divided equally between inelastic and elastic processes with 
weight $\sigma_0/4$ for $|\omega_{p\pm}|\ll \omega \ll W$, similarly to the channel isotropic two-channel Kondo 
model at $\omega=0$, which is a prototype model for non-Fermi liquid behaviour. 
However, 
the classification of impurity states as Fermi or non-Fermi liquid becomes 
inappropriate in the present case, since scattering is caused by bosonic, and 
not fermionic modes. 
%This similarity is further corroborated by the comparison of the spectral functions (of phonons in our 
%case\cite{doraphonon} and of the spins for the Kondo\cite{emery}). Both 
%contains two Lorentzians, one of which shrinks to a Dirac-delta function 
%at $\Gamma_2$ for the phonons 
%or at the channel isotropic limit of the Kondo model\cite{nersesyan}.
Since the transition occurs at $g\rightarrow\infty$ ($\Gamma\rightarrow\Gamma_2$), the crossing of 
this critical point is impossible.

Fig. \ref{sigmaGall} shows %s. \ref{sigmaG1} and \ref{sigmaG2} show 
the total, elastic and inelastic scattering rates as 
a function of frequency 
for several $\Gamma$. At zero frequency, all rates vanish, since the electron Green's function is pinned to its 
non-interacting value due to some generalized Fermi liquid relations\cite{hewson,cornaglia1}.
For small couplings and $\omega<$Re$\omega_{p\pm}$, the ground state contains no phonons but a filled Fermi sea, 
thus electrons scatter off elastically. By exceeding this threshold (Re$\omega_{p\pm}$), the total scattering is mainly 
given by the 
inelastic 
cross section, which develops steps at 
at $n\times|\omega_{p\pm}|$ (with $n$ integer), stemming from $n$-phonon excitations.
At $n=1$, this is also predicted by perturbation theory\cite{engelsberg} as 
$\sigma_{tot}\approx\sigma_{inel}=\sigma_0\Gamma\pi\Theta(\omega-\omega_0)/\omega_0$.
Higher steps occur with smaller weight  in alternating fashion.
At the same time, the elastic rate exhibits a small cusp at $\omega_0$ due to the 
logarithmic singularity of the real part of electron self energy\cite{engelsberg}.

For larger couplings, these sharp steps become smeared and turn into oscillations due
to the finite lifetime of the phonons. These oscillation are not of Friedel type, 
corresponding to the interference of incoming and scattered waves, but arise from multiphonon excitations.
The elastic scattering rate grows progressively, and 
becomes the dominant scattering mechanism for 1 and 2 phonon excitations.

By approaching the critical coupling $\Gamma_2$, the inelastic scattering reaches 
its maximum, since the vanishing phonon 
energy, $\omega_{p-}$, makes arbitrary multi-phonon processes possible.
We estimate\cite{park,weig} $\omega_0$ to fall into the 0.5-50~K range, which compares well to the typical temperatures of weak localization experiments.
These results at $T=0$ are in marked contrast to those found for Kondo 
impurities\cite{borda,micklitz,bauerle}.

In actual experiments (on e.g. Ag or Au samples with Fe impurities), 
the dephasing time is measured at finite temperatures, and 
can be decomposed as $1/\tau_\phi=1/\tau_{e-e}+1/\tau_{e-ph}+1/\tau_{imp}$.
In the high temperature limit, it is dominated by the bulk electron-electron 
($1/\tau_{e-e}\sim T^{2/3}$) and electron-phonon ($1/\tau_{e-ph}\sim T^{3}$) 
interactions, and is accounted for by the Altshuler-Aronov-Khmelnitzky theory\cite{AAK}. 
At lower temperatures, 
the impurity contribution becomes important ($1/\tau_{imp}\sim n_{imp}\sigma_{inel}$), and scales with the impurity 
concentration $n_{imp}$.
Magnetic impurities account successfully for this latest contribution down to 0.1$T_K$ (with $T_K$ the Kondo 
temperature).

Our calculation on the local oscillator model can be extended to finite temperatures.
Conformal invariance can only be used for times much bigger\cite{nersesyan} than $1/\omega_0$, which translates to 
$T\ll \omega_0$, leading to $T^2$ corrections. The same conclusion borns out from the 
generalized Fermi liquid relations\cite{hewson,cornaglia1}.
In the 4Re$\omega_{p\pm} \lesssim T\ll W$ range, the inelastic scattering rate reaches its 
maximal value, $\sigma_{inel}=\sigma_{el}=\sigma_0/4$, since at high temperatures, 
the phonon state is highly populated, facilitating inelastic scattering. This behaviour 
is similar to the $T=0$ case close to the softening $\Gamma\rightarrow\Gamma_2$, where 
the vanishing phonon excitation frequency favours inelastic processes.
In between these two regions ($T\sim \omega_0$), a smooth crossover takes place.
%Therefore, this picture is reminiscent to that of a Kondo impurity, which follows a Fermi %liquid behaviour below the Kondo scale, and develops a broad plateau above it. 
%In comparison
%with experimental results, these are the necessary ingredients to explain the measured %dephasing time at low temperatures\cite{bauerle}. Therefore, local vibrational modes can %account successively to
%the measured dephasing time. 
%These aspects look particularly promising upon realizing, that t

The Kondo picture starts to deviate from the experimental 
data below the Kondo temperature\cite{bauerle,alzoubi,mallet}. This mismatch is cured by 
introducing a very small amount of magnetic impurities with a very low Kondo temperature\cite{bauerle} or 
by adding a constant background\cite{mallet}.
Instead, a flat, temperature independent dephasing time (with $\omega_0\ll T_K\ll W$) follows naturally from 
local 
vibrational modes, as was demonstrated above, and serves as an obvious explanation
for the measured behaviour.
Therefore, local vibrational modes can
 account successively for the measured dephasing time in weak localization experiments.

The second crucial difference with respect to Kondo impurities arises from the magnetic field 
insensitivity of the phonons, and requires further experiments to reveal the differences\cite{huang}.

%, and we expect similar behaviour as a function of 
%$T$: the cross sections at zero frequency vanish at $T=0$ and saturate to a finite value for %$\omega_0\ll T\ll W$.
%A possible extension to the case of a correlated conduction band might also be possible, %since no refermionization is required to diagonalize Eq. \eqref{hamilton}, unlike in the
% Kondo model\cite{nersesyan}.

In conclusion, we have demonstrated through an exact solution of the single impurity Holstein or Yu-Anderson model, 
that local vibrational modes can have a strong impact on the dephasing time of electrons.
The inelastic scattering rate exhibits strong oscillations at frequencies comparable to the phonon excitation energy, 
and then saturates to a finite, coupling dependent value. At the extreme strong coupling limit, close to the complete 
softening of the phonons, the $s$-matrix vanishes and the inelastic cross section reaches its maximal value. 
This phonon mediated scattering mechanism is expected to be rather insensitive to the applied magnetic field, in 
contrast to Kondo-type impurities, and can contribute to the dephasing time in certain 
alloys containing dynamical defects\cite{park,huang}.

%\begin{acknowledgments}
Useful and illuminating discussions with Peter Fulde and Sergej Flach are gratefully acknowledged.
This work was supported by the Hungarian
Scientific Research Fund under grant number K72613.
%\end{acknowledgments}

\bibliographystyle{apsrev}
\bibliography{wboson}
\end{document}